\documentclass[english,a4paper]{article}
\usepackage[T1]{fontenc}
\usepackage[cp1250]{inputenc}
\usepackage{amssymb}
\usepackage{hyperref}
\usepackage{amsfonts}
\usepackage{graphicx}

\begin{document}
\title{Time-optimal monotonic convergent algorithms for the control of quantum systems}

\author{M. Lapert, J. Salomon\footnote{CEREMADE, Universit\'e Paris Dauphine, Place du
Mar\'echal De Lattre De Tassigny, 75775 Paris Cedex 16, France}
and D. Sugny\footnote{Laboratoire Interdisciplinaire Carnot de
Bourgogne (ICB), UMR 5209 CNRS-Universit\'e de Bourgogne, 9 Av. A.
Savary, BP 47 870, F-21078 DIJON Cedex, FRANCE,
dominique.sugny@u-bourgogne.fr}}

\maketitle

\begin{abstract}
We present a new formulation of monotonically convergent
algorithms which allows to optimize both the control duration and
the field fluence. A standard algorithm designs a control field of
fixed duration which both brings the system close to the target
state and minimizes its fluence, whereas here we include in
addition the optimization of the duration in the cost functional.
We apply this new algorithm to the control of spin systems in
Nuclear Magnetic Resonance. We show how to implement CNOT gates in
systems of two and four coupled spins.
\end{abstract}
\section{Introduction}
The optimal control of quantum systems is a long-standing goal
\cite{warren,rabitz,daniel} which remains very attractive both
from a practical and a fundamental point of views
\cite{shapiro,rice,tannorbook}. By finding the optimal balance
between the minimization of the distance to the target state and
the minimization of the energy of the control field, the optimal
solution allows to bring the system close to the target state
while avoiding parasitic phenomena due to a too large fluence. In
this context, the control duration is also a crucial parameter
which has to be taken into account in the optimization process.
For instance, a too long duration could be problematic if other
concurrent physical or chemical processes with the same time scale
occur during the control. This question is particularly
interesting in quantum computing where coherence
 has to be preserved \cite{nielsen}. Since a control field cannot generally fully compensate the dissipation effects \cite{altafini}, a too long interaction of the system with the environment can destroy
 its coherence and the quantum superposition or the entanglement produced
 by the control.

Solving time-optimal control problems remains however a
challenging task. One way is to use the
 Pontryagin maximum principle and geometric optimal control theory \cite{jurdjevic,bonnard}.
 However, such techniques can only be applied for the moment to
 small dimensional quantum problems with very few energy levels
 \cite{boscain,sugny1,carlini,khaneja,lapertglaser}. On the other hand,
 monotonically convergent algorithms are an efficient way to solve
 optimal control problems and have been widely used in the control
 of chemical and physical processes since the pioneering papers by
 Tannor et al. \cite{tannor} and Rabitz et al. \cite{rabitz1}
 which were based on the work of Krotov \cite{krotov}. This approach can be
 applied to very different and large quantum systems (See e.g. \cite{ohtsuki1,ohtsuki2,werschnik,salomon,bifurcating1}) and to a variety of
 non-standard situations such as the nonlinear interaction between the system and the control field \cite{lapert1,nakagami} or to take into account spectral constraints on the optimal
 solution \cite{gollub,lapert2}. Up to now, however, these algorithms have generally been used
 with a cost penalizing the field fluence and a fixed control
duration. By construction of these algorithms, a formulation in
terms of a time-optimal control, i.e. with a duration which is not
fixed, is a very difficult question since these methods imply the
backward propagation of the adjoint state from the final time of
the control. A first possibility to reduce the time of control
consists of penalizing the intensity of the control field in a
neighborhood of the final time. However, this technique strongly
depends on the penalization, and requires adjustments that have to
be done by the operator \cite{salomon}. A time-optimal or free
time optimal control algorithm was proposed in \cite{yamashita},
but this method differs from our approach in the sense that a
second Lagrange multiplier on the control duration (in addition of
the adjoint state) is added. This leads to a more complicated
algorithm than the one proposed below.

We present in this paper a new formulation of monotonically
convergent algorithms with a cost penalizing both the field
fluence and the control duration, which allows us to find the best
compromise between these two parameters. Using a rescaling of
time, we first rewrite the optimal equations on a fixed time
interval independent on the control duration $T$, which appears as
a new parameter in the time-dependent Schr\"odinger equation. We
 then consider a monotonic iterative algorithm whose each step is
 decomposed into two substeps consisting in an optimization of the
 energy of the field with $T$ fixed and an optimization of the
 time $T$ with a fixed control field. The first substep is done by
 a standard monotonically convergent algorithm, while a gradient or another discrete optimization
 procedure is used for the second substep. We impose that each
 substep increases the cost functional leading thus to a monotonic
 algorithm. This algorithm is described in a very general
 way and can therefore be applied to any problem of quantum
 control.

To test the efficiency of this approach, we consider the control
of spin systems \cite{spin}, and in particular the implementation
of quantum gates in such systems. Different technologies have been
developed so far to exploit the powerful of quantum computing. One
of the most promising solution is Nuclear Magnetic Resonance (NMR)
\cite{vander}. The control technology developed over the past
fifty years allows the use of sophisticated control fields and
permits the implementation of complex quantum algorithms such as
the Deutsch-Jozsa and the Grover ones \cite{chuang}. NMR is
therefore an ideal testbed to experiment new ideas in quantum
control. In this paper, we show how to implement two and four
qubits CNOT gates. These different numerical computations allow to
extract the main properties of our algorithm and to highlight the
differences with respect to a standard approach.

The paper is organized as follows. In Sec. \ref{sec2}, we describe
the new monotonically algorithm for pure state quantum systems. The proof of its monotonic character is established.
Section \ref{sec3} is devoted to the application of this approach
in a two and four spin systems in order to implement, in optimized time,
CNOT gates. We conclude in Sec.
\ref{sec4}.
\section{Time-optimal control algorithms}\label{sec2}
We present in this section the algorithm in a general setting for pure quantum
states. The formalism can be straightforwardly extended to
mixed-state quantum systems \cite{ohtsuki2} or to the control of
evolution operators \cite{kosloff}. It is this latter generalization that
will be used in Sec. \ref{sec3} for the implementation of quantum gates. We consider the maximization
of the projection onto a target state, but the algorithm could be
equivalently defined for maximizing the expectation value of a
given observable.
\subsection{Methodology}\label{sec2a}
Let $|\phi_0\rangle$ and $|\phi_f\rangle$ be the initial and
target states of the dynamics. We consider the time-optimal control
problem of maximization of the cost functional $J_T(E_T)$ over the
control duration $T$ and the control field $E_T$. Note that the subscript $T$ is added in this paper to any quantity depending upon this time. The functional $J_T$ is defined by
\begin{equation}\label{eq1}
J_T(E_T)=2\Re[\langle \psi_T(T)|\psi_f\rangle]-\alpha
\int_0^TE_T^2(t)dt
\end{equation}
where $\alpha$ is a positive parameter which weights the relative
importance of the energy of the control field with respect to the
projection onto the target state. $\Re [\cdot]$ is the real part
of a complex number. The state $|\psi_T(t)\rangle$ of the system
satisfies the time-dependent Schr\"odinger equation which is
written in units such that $\hbar=1$:
\begin{equation}\label{eq2}
i\frac{\partial}{\partial
t}|\psi_T(t)\rangle=(H_0+E_T(t)H_1)|\psi_T(t)\rangle
\end{equation}
with as initial condition $|\psi_T(0)\rangle=|\phi_0\rangle$. The
Hamiltonian $H_0$ is the field-free Hamiltonian and the operator
$H_1$ describes the interaction between the system and the control
field, which is assumed to be linear.

The first step of the method is to define a fixed time interval,
for instance $[0,1]$. We consider for that purpose the time
rescaling $s=t/T$. Introducing $|\psi(s)\rangle=|\psi_T(s\cdot
T)\rangle$ and $E(s)=E_T(s\cdot T)$, we obtain from Eq.
(\ref{eq2}) that
\begin{equation}\label{eq3}
i\frac{\partial}{\partial
s}|\psi(s)\rangle=T(H_0+E(s)H_1)|\psi(s)\rangle
\end{equation}
with the initial condition $|\psi(0)\rangle=|\phi_0\rangle$. The
cost functional is also changed by the time rescaling and becomes
\begin{equation}\label{eq4}
J(E)=2\Re[\langle \psi(1)|\psi_f\rangle]-\alpha T\int_0^1E^2(s)ds.
\end{equation}
The new optimal control problem consists now in maximizing the
cost functional $J$ with respect to the control field $E$ and the
time $T$ which plays here the role of a parameter. The control
duration is fixed to 1.
\subsection{Monotonically algorithm}\label{sec2b}
The algorithm is decomposed into two substeps. We alternatively optimize the functional $J$ with respect to the control field $E$
by a standard monotonic algorithm and with respect to the duration $T$ by a discrete procedure such as a gradient method.
We prove that the cost increases at each step of the
algorithm.
\paragraph{Optimization of the control field.}
We consider the triplets $(|\psi(t)\rangle,E(t),T)$ and
$(|\tilde{\psi}(t)\rangle,\tilde{E}(t),T)$ corresponding to the
initial and final states of this substep of the algorithm. The
variation of the cost is given by:
\begin{eqnarray*}\label{eq5}
\Delta J&=& J(\tilde{E})-J(E)\\
&=&2\Re[\langle
\tilde{\psi}(1)-\psi(1)|\psi_f\rangle]-\alpha
T\int_0^1(\tilde{E}^2(s)-E^2(s)ds.
\end{eqnarray*}
We introduce the adjoint state $|\chi(t)\rangle$ which satisfies
\begin{equation}\label{eq6}
i\frac{\partial}{\partial
s}|\chi(s)\rangle=T(H_0+E(s)H_1)|\chi(s)\rangle
\end{equation}
with the final condition $|\chi(1)\rangle=|\phi_f\rangle$. We then
have
\begin{equation}\label{eq7}
\Re[\langle \tilde{\psi}(1)-\psi(1)|\phi_f\rangle]=\Re[\langle
\tilde{\psi}(1)-\psi(1)|\chi(1)\rangle]
\end{equation}
which can be transformed into
\begin{equation}\label{eq7}
\Re[\langle
\tilde{\psi}(1)-\psi(1)|\phi_f\rangle]=\Re\big[\int_0^1ds[\langle
\frac{\partial}{\partial s}\chi|\tilde{\psi}-\psi\rangle+ \langle
\chi|\frac{\partial}{\partial s}(\tilde{\psi}-\psi)\rangle]\big].
\end{equation}
Using Eqs. (\ref{eq3}) and (\ref{eq6}), one deduces that
\begin{equation}\label{eq8}
\Re[\langle
\tilde{\psi}(1)-\psi(1)|\phi_f\rangle]=2T\Im[\int_0^1ds\langle
\chi|H_1 (E-\tilde{E})|\tilde{\psi}\rangle] .
\end{equation}
One finally arrives to
\begin{equation}\label{eq9}
\Delta J=\alpha T\int_0^1
ds(E-\tilde{E})(E+\tilde{E}-\frac{2}{\alpha}\Im[\langle
\chi|H_1|\tilde{\psi}\rangle]).
\end{equation}
Knowing $E(s)$, the choice $\tilde{E}=\Im[\langle
\chi|H_1|\tilde{\psi}\rangle]/\alpha$ ensures that $\Delta J\geq
0$ for this substep. This part of the algorithm can be summarized
as follows. Starting from the quadruplet
$(|\psi(s)\rangle,|\chi(s)\rangle,E(s),T)$, we construct the
quadruplet of the next sub-iteration by propagating backward the
adjoint state $|\tilde{\chi}(s)\rangle$ from $|\phi_f\rangle$ with
the field $E(s)$. We then propagate forward the state
$|\tilde{\psi}(s)\rangle$ from $|\phi_0\rangle$ with the field
$\tilde{E}(s)$ which is computed at the same time by the relation
$\tilde{E}(s)=\Im[\langle
\tilde{\chi}(s)|H_1|\tilde{\psi}(s)\rangle]/\alpha$.
\paragraph{Optimization of the control duration.} At this stage of the algorithm, we
consider the triplets $(|\psi(s)\rangle,E(s),T)$ and
$(|\tilde{\psi}(s)\rangle,E(s),\tilde{T})$. We recall that the
state $|\tilde{\psi}(s)\rangle$ satisfies
\begin{equation}\label{eq10}
i\frac{\partial}{\partial
s}|\tilde{\psi}(s)\rangle=\tilde{T}(H_0+E(s)H_1)|\tilde{\psi}(s)\rangle.
\end{equation}
We compute the variation of the cost functional $\Delta J$ which
is equal to:
\begin{equation}\label{eq11}
\Delta J=2\Re[\langle
\tilde{\psi}(1)-\psi(1)|\psi_f\rangle]-\alpha
(\tilde{T}-T)\int_0^1E^2(s)ds.
\end{equation}
Introducing the adjoint state $|\chi(s)\rangle$ whose dynamics is
governed by Eq. (\ref{eq6}), one obtains after similar computations as for the previous case
that
\begin{equation}\label{eq13}
\Delta J=\alpha(T-\tilde{T})\int_0^1ds
E(s)(-\frac{2}{\alpha}\Im[\langle
\chi|H_1|\tilde{\psi}\rangle]+E(s)).
\end{equation}
The parameter $\tilde{T}$ has to be chosen such that $\Delta J\geq
0$. A solution consists in using a gradient method by noting that
\begin{equation}\label{eq14}
\nabla_T J=-\alpha\int_0^1ds E(s)(\frac{2}{\alpha}\Im[\langle
\chi|\mu|\psi\rangle]+E(s)).
\end{equation}
We define the new time $\tilde{T}$ from the preceding one as:
\begin{equation}\label{eq15}
\tilde{T}=T-r\nabla_TJ(E(s),T),
\end{equation}
where $r$ is a small real parameter. We choose numerically $r$
small enough to ensure the monotonicity of the cost functional. The computation of the optimal value of $r$ requires however several new
propagations to determine the cost $J$ since the evolution of $|\tilde{\psi}\rangle$ (needed to calculate $\langle \tilde{\psi}(1)|\psi_f\rangle$) depends on the value of $\tilde{T}$ (see Eq. (\ref{eq10})). Other methods ensuring the monotonic behavior of the cost can be used for this substep as the following procedure. In this approach, we define the new duration $\tilde{T}_k$ as a function of the old duration $T_k$ as follows:
\[
\tilde{T}_k=(1+a)T_k
\]
where $a$ is a small positive or negative parameter. Practically,
we can choose e.g. $a=\pm 10^{-3}$, but this value can also be
adjusted during the computation. This  leads to two new costs
$\tilde{J}_k^+$ and $\tilde{J}_k^-$. The final time at step $k$ is
the time associated to the maximum value between $\tilde{J}_k^+$,
$\tilde{J}_k$ and $\tilde{J}_k^-$. This method has the advantage
over the gradient approach to limit at each step the number of
propagations of Eq. (\ref{eq3}) to 2. This point can be
interesting when very heavy computations are considered. This
systematic procedure has been used in the numerical examples of
Sec. \ref{sec3}. In particular cases, we have checked that the
gradient and this systematic approach give equivalent results.
Note that a faster algorithm can be designed by not following a
strict alternation between the two optimization procedures. In
other words, this means that the control field can be optimized
several times between each optimization of the time parameter.
However, such a method requires adjustments and a more involved
study which are out the scope of this paper.
\section{Control of spins systems}\label{sec3}
\subsection{Description of the model}\label{sec3a}
The principles of control in NMR are detailed in different books
and review articles. Here we only give a brief account needed to
introduce the model used \cite{spin}. We consider the control of a
system of coupled spins by different magnetic fields acting as
local controls on each spin. This means that each field only
controls one spin and does not interact with the others, i.e. the
spins are assumed to be selectively addressable. This hypothesis
has also the advantage to render the system completely
controllable. Similar models have been used in numerical studies
analyzing the realization of quantum algorithms in NMR (see, e.g.,
Ref. \cite{schulte}).

We introduce a system of $n$ coupled spins whose evolution is
described by the following Hamiltonian:
\[
H=H_0+\sum_{j=1}^n(u_{jx}H_{jx}+u_{jy}H_{jy}),
\]
the couplings being given by:
\[
H_{jx}=\sigma_{jx},~H_{jy}=\sigma_{jy}
\]
where the operators ($\sigma_{jx}$, $\sigma_{jy}$) are Pauli matrices which only act on the $j$th- spin. We assume that the free evolution Hamiltonian $H_0$ is associated to the topology of a chain of coupled spins with only nearest-neighbor interactions. The corresponding Hamiltonian is given by:
\[
H_0=\sum_{j=1}^{n-1}\sigma_{jz}\otimes \sigma_{j+1;~z},
\]
where the approximation is valid in heteronuclear spin systems if
the coupling strength between the spins is small with respect to
the frequency shifts \cite{spin}. The coupling parameter between
the spins is taken to be uniform and equal to 1. The different
equations being linear, other couplings could be considered from a
standard rescaling of the time and of the amplitude of the control
fields. Note that the algorithm could also be used with different
couplings between the spins.
\subsection{Optimal implementation of a CNOT gate}\label{sec3b}
Our goal is to apply the time-optimal control algorithm to
implement a $C^{n-1}NOT$ gate (Controlled-Not) in a system of $n$
qubits with $n=2$ or 4. A $C^{n-1}NOT$ gate is a gate in which the
target qubit flips if and only if the $(n-1)$ control qubits are
equal to 1. For $n=2$, the $CNOT$ transformation is represented by
the unitary operator $U_{CNOT}$ which can be written as:
\begin{eqnarray*}
U_{CNOT}=\left(
\begin{array}{cccc}
1 & 0 & 0 & 0\\
0 & 1 & 0 & 0\\
0 & 0 & 0 & 1\\
0 & 0 & 1 & 0
\end{array}\right).
\end{eqnarray*}
The logical states involved in a $C^{n-1}NOT$ gate can be mapped
onto the spin states in different ways. A straightforward and
natural way used in this paper is to encode the first qubit in the
first spin, the second qubit in the second spin, and so on if more
than two spins are considered.

To implement quantum gates, we formulate the control problem in
terms of evolution operators $U(t)$. This means that the objective
of the control is to reach the target state $U_{CNOT}$, while
optimizing the control duration and the energy of the fields. The
time-optimal monotonically convergent algorithm for evolution
operators can be sketched along the same way as for the wave
function case of Sec. \ref{sec2}. This algorithm can be obtained
by replacing the wave function $|\psi_T(t)\rangle$ by $U_T(t)$.
The corresponding cost functional is given by:
\begin{equation}
J(E)=2\Re \big[\textrm{Tr}[U_{CNOT}U_T(T)]\big]-\alpha \int_0^T E_T^2(t)dt.
\end{equation}
The evolution operator $U_T(t)$ satisfies the Schr\"odinger equation
\begin{equation}
i\frac{\partial U_T(t)}{\partial t}=(H_0+E_T(t)H_1)U_T(t)
\end{equation}
and the scalar product $\langle \psi(t)|\chi(t)\rangle$ is
replaced by $\textrm{Tr}[U\dag V]$ where $V(t)$ is the adjoint
propagator. Note that, in this case, $2^n$ fields are
simultaneously optimized. The efficiency of the process is
measured by the projection $P=\frac{1}{2^n}\Re
\big[\textrm{Tr}[U_{CNOT}U_T(T)]\big]$.

From a numerical point of view, two different parameters,
$\alpha_0$ and $E_0(t)$, have to be adjusted when using this
algorithm. These parameters do not play the same role since
$\alpha$ is a parameter characteristic of the algorithm, while
$E_0(t)$ is the initial field (the same for all the fields) used
to initiate the optimization process. More precisely, we assume
that the parameter $\alpha$ depends on time and can be written as
$\alpha(t)=\alpha_0\sin^2(\pi t/T)$ where $\alpha_0$ is a
constant. This switching function is introduced to provide a
smooth on and off switch of the field \cite{korff}. In order to
not enforce the algorithm to follow a given pathway, we consider
that the initial trial field $E_0(t)$ is zero over a given
duration $T_0$. The dependance of the final solution on the two
parameters $T_0$ and $\alpha_0$ will be analyzed in Sec.
\ref{secnum}.

\subsection{Numerical results}\label{secnum}
\begin{figure}
\centering
\includegraphics[width=0.4\textwidth]{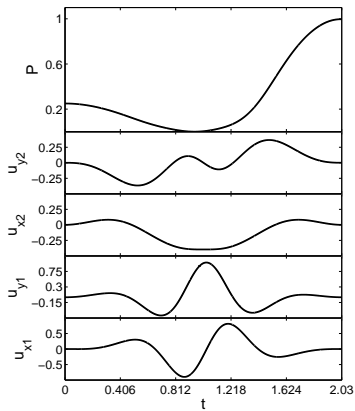}
\caption{\label{fig1} Optimization of the CNOT gate: Evolution of
the probability $P$ (top) and of the corresponding optimal fields
acting on the first or second spin (bottom). Numerical values are
taken to be $\alpha_0=0.08$ and $T_0=0.5$. The final probability
is $P=0.9964$.}
\end{figure}

\begin{figure}
\centering
\includegraphics[width=0.4\textwidth]{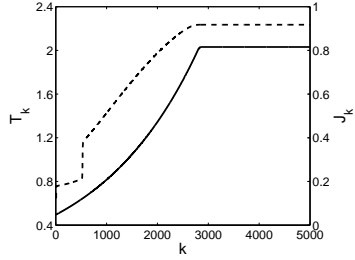}
\caption{\label{fig2} Optimization of the CNOT gate: Evolution of
the cost $J_k$ (dashed line) and of the time $T_k$ (solid line) as
a function of the number of iterations $k$ of the algorithm. The
same parameters ($\alpha_0$ and $T_0$) as in Fig. \ref{fig1} have
been used.}
\end{figure}

\begin{figure}
\includegraphics[width=0.4\textwidth]{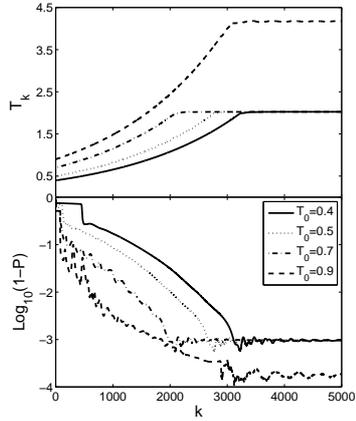}
\caption{\label{fig3} Optimization of the CNOT gate: (top)
Evolution of the time $T_k$ for different initial times $T_0$ as a
function of the number $k$ of iterations. For $T_0\leq 0.7$, the
algorithm converges towards the same optimal duration close to the
value 2. The parameter $\alpha_0$ is taken to be 0.08. (bottom)
Same as before but for the probability density $P$. A better
efficiency is reached for $T_0=0.9$.}
\end{figure}

We first analyze the computational results for a system of two
spins. Figure \ref{fig1} displays the optimal solution computed by
the algorithm for the values of parameters $T_0=0.5$ and
$\alpha_0=0.08$. The parameter $a$ which describes the evolution
of the control duration at each step of the algorithm is taken to
be $5\times 10^{-4}$. Other values of $a$ have been used leading
either to worse results or to a slower convergence of the
algorithm. As can be seen in Fig. \ref{fig1}, the evolution of the
optimal control fields and of the probability is rather smooth
with no rapid oscillation. A very good efficiency larger than 0.99
has been reached in 5000 iterations with a final duration of the
order of $T=2.035$. Note that a standard monotonically algorithm
with this total duration leads to a solution very close to the
ones obtained with this new algorithm. More precisely, for the
standard algorithm, we have obtained a projection $P$ larger than
0.999 for a control duration such that $1.9<T<2.4$. This
computation shows that the time-optimal control algorithm has
found the best compromise between the duration, the minimization
of the distance to the target state and the energy of the field.
As could be expected, the modification of the control duration
slows down the convergence of the algorithm since a projection
larger than 0.99 is obtained respectively after 2512 and 700
iterations for the new and standard methods, respectively. The
monotonic behavior of the algorithm can be checked in Fig.
\ref{fig2} together with the evolution of the duration $T_k$. As
for the cost $J_k$, one sees that this parameter presents a rapid
increase for the first 3000 iterations and then an approximatively
constant behavior.
A crucial property that this algorithm must satisfy (at least locally) is the independence of the final solution with respect to the value of $T_0$, i.e. of the starting guess used to initiate the algorithm. This point is illustrated in Fig. \ref{fig3} where two attraction points for the sequence $(T_k)$ have been found when the time $T_0$ varies. We numerically determine the two basins of attraction and we found a boundary of the order of $T_0\simeq 0.75$. Other attraction points exist for larger initial values of the control duration $T_0$ which are not represented in Fig. \ref{fig3}. Note that this attraction point characterizes not only the final control duration but also the final control fields and the final probability density as can been checked in Fig. \ref{fig3}.
We also see in this figure that a better efficiency is reached for
$T_0=0.9$ with a longer and lower energetic optimal solution. This
point stresses the role of the control duration in the accuracy of
the computation.

\begin{figure}
\includegraphics[width=0.4\textwidth]{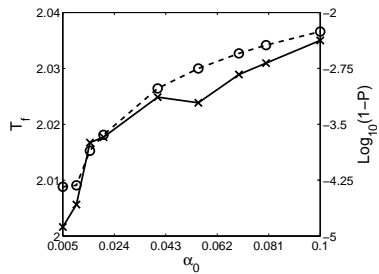}
\caption{\label{fig4} Optimization of the CNOT gate: Evolution of
$T_{f}$ (open circle) and $P$ (cross) as a function of the
parameter $\alpha_0$ for $T_0=0.5$. The solid and dashed lines are
just to guide the lecture.}
\end{figure}

In Fig. \ref{fig4}, we study the evolution of the final time $T_f$ and of the probability density $P$ as a function of the parameter $\alpha_0$. We observe that $P$ increases and $T_f$ decreases as the parameter $\alpha_0$ decreases. As could be expected, the smaller $\alpha_0$ is, the more energetic the optimal solution is since $\alpha_0$ controls the relative weight of the pulse energy in the cost $J$. With a more energetic optimal solution, the algorithm can find an optimal solution with a lower duration and a better efficiency.


\begin{figure}
\includegraphics[width=0.4\textwidth]{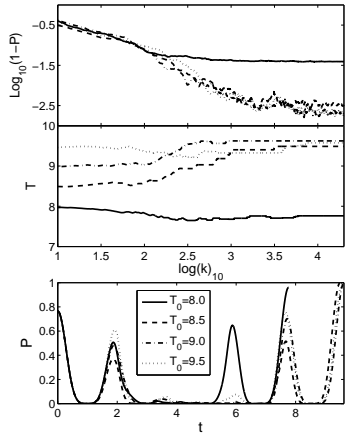}
\caption{\label{fig5} Optimization of the C$^3$NOT gate:
(top-middle) Evolution of the probability density $P$ and of the
control duration $T_k$ as a function of the number of iterations
for different initial durations $T_0$. (bottom) Time-evolution of
the probability density $P$ for the different optimal solutions.
The parameter $\alpha_0$ is taken to be $\alpha_0=0.01$.}
\end{figure}

We extend these numerical results to the case of a four-spin
system and a C$^3$NOT gate. Due to the complexity of this gate, a
larger duration and a larger number of iterations are required to
reach a sufficient efficiency. The parameter $a$ is taken to be
$5\times 10^{-4}$. As for the two-spin case, we find two possible
optimal solutions according to the value of $T_0$ which are
displayed in Fig. \ref{fig5}. The time evolution of the
probability density shows that the structure of these two
solutions is very similar even if the two final durations are
different.
\section{Conclusion and perspectives}\label{sec4}
This work deals with the time-optimal control of spin systems in
NMR. We propose a monotonically convergent algorithm which both
optimizes the control duration and the energy of the field. We
show that the change of the duration at each iteration of the
algorithm leads to a more flexible algorithm and thus allows a
better convergence with respect to a standard version of such
algorithms. This method has the advantage of simplicity and
general applicability whatever the quantum optimal control problem
considered. We have finally demonstrated the possibility of
implementing quantum gates from the control fields computed by
this algorithm. Since there exists no unique optimal solution, we
have shown that we can select the control fields by changing the
initial duration of the control.

\end{document}